\documentclass[pra,aps,reprint,showpacs,amsmath,superscriptaddress]{revtex4-1}

\usepackage{bm}
\usepackage{graphicx}
\usepackage{dsfont}

\usepackage{bm}
\usepackage{mathrsfs}

\usepackage{framed}

\usepackage[all]{xy}
\usepackage{graphicx}
\usepackage{framed}
\usepackage{comment}
\usepackage{here}                    % manages figures
\usepackage{latexsym}		     % to get LASY symbols
\usepackage{amsmath}     
\usepackage{amsfonts}  
\usepackage{amssymb}
\usepackage{amsthm}
\usepackage{dcolumn}
\usepackage{color}
\usepackage{mathrsfs}

\renewcommand{\phi}{\varphi}
\renewcommand{\>}{\rangle}
\newcommand{\<}{\langle}

\newcommand{\ket}[1]{|#1\>}
\newcommand{\bra}[1]{\<#1|}

\newcommand{\be}{\begin{equation}}
\newcommand{\ee}{\end{equation}}
\newcommand{\bea}{\begin{eqnarray}}
\newcommand{\eea}{\end{eqnarray}}

\newcommand{\Int}{\mathbb{Z}}

\newcommand{\aver}[1]{\langle #1 \rangle}

\renewcommand{\phi}{\varphi}

\begin{document}

\title{Continuous-variable quantum authentication of physical unclonable keys: Security  against an emulation attack}

\author{Georgios M. Nikolopoulos}
\email{nikolg@iesl.forth.gr}

\affiliation{Institute of Electronic Structure \& Laser, 
FORTH, P.O. Box 1385, GR-70013 Heraklion, Greece}

\date{\today}

\begin{abstract}
We consider a recently proposed entity authentication protocol, in which a physical unclonable key is interrogated by random coherent states of light, and the quadratures of the scattered light are analysed by means of a coarse-grained  homodyne detection. 
We derive a sufficient condition for the protocol to be secure against an emulation attack, in which an adversary knows the challenge-response properties of the key, and moreover he can access the challenges during  the verification. The security analysis relies on Holevo's bound and Fano's inequality, and suggests that the protocol  is secure against the emulation attack for a broad range of physical parameters that are within reach of today's technology.
\end{abstract}


\maketitle

%%%%%%%%%%%%%%%%%%%%%%%%%
%             SECTION  I
%%%%%%%%%%%%%%%%%%%%%%%%%

\section{Introduction} 
\label{sec1}

The development of entity authentication (identification) protocols (EAPs), which are robust against 
cloning and other impersonation strategies, is an outstanding task in modern cryptography
\cite{handbook,handbook2}. Currently, optical physical unclonable keys (PUKs) are considered to be the most promising candidates for the development of highly robust EAPs \cite{Horstmeyer15,Tuyls05,Pappu02, PappuPhD,Buch05,Iesl16,Goorden14,Skoric13,NikDiaSciRep17}. Typically, optical PUKs are materialized by multiple-scattering optical media, where disorder is introduced during  fabrication. Hence, faithful cloning of an optical PUK requires  the exact positioning (on a nanometer scale) of millions of scatterers with the exact size and shape, which is considered to be a formidable challenge not only for current, but for future technologies as well \cite{Pappu02, PappuPhD,Goorden14}.   

Optical PUK-based authentication protocols
\cite{Horstmeyer15,Tuyls05,Pappu02,PappuPhD,Buch05,Iesl16,Goorden14,Skoric13,NikDiaSciRep17} 
rely on two-stage challenge-response mechanisms,  
analogous to the ones used in everyday transactions with smart 
cards (e.g., credit cards), at automatic teller machines, and point of sale terminals 
\cite{handbook,handbook2}.  
The {\em enrolment stage} is performed only once by the manufacturer (or the authority that issues 
and distributes the PUKs), well before the PUK is given to a user. It aims at the faithful characterization 
of the PUK with respect to its optical responses to a finite set of random challenges, that is to light 
pulses (probes) with random parameters.  The  set of challenge-response pairs (CRPs) is stored at a server, and the PUK is given to a user, together with 
a personal identification number (PIN), which is also saved at the same server. The {\em verification stage} takes place each time the holder of 
the PUK  has to be authenticated to the system, it is performed by a legitimate verifier, who is connected to the server over a secure and authenticated classical channel, and involves 
two steps. First, the user inserts his PUK in a verification device and types in his secret PIN. In this way 
the user is verified to the PUK, thereby preventing someone who has stolen the PUK  and does not know the PIN,  to 
pass himself as the user. The PIN is sent to the server over the classical 
channel, and if it is valid the verifier has to decide whether the PUK is authentic or not.  To this end, the server sends to the verifier (over the classical channel), a moderate number of challenges, chosen at random from the recorded challenges in the  database. 
The verifier interrogates sequentially the PUK with each one of the challenges, and the corresponding responses are sent back to the server, where they are compared to those in the database.  The server decides upon the acceptance (rejection) of the PUK, based on some algorithm that takes into account the deviations from the expected responses.  
The number of challenges that are used in the verification stage depends on the EAP under consideration, but in any case it should be such that the verification stage takes place within a reasonable amount of time, say few seconds (see also related discussion in Sec. \ref{sec2b}).  

\begin{figure*}
\includegraphics[scale=0.6]{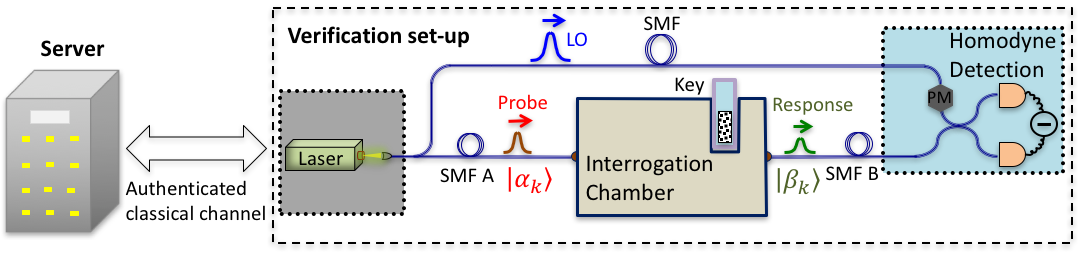}
\caption{(Color online) A schematic representation of the PUK-based optical EAP under consideration. 
}
\label{fig1}
\end{figure*}

Most of the proposed optical EAPs that rely on PUKs, exploit the speckle of the scattered light 
and pertain to classical challenges \cite{Horstmeyer15,Tuyls05,Pappu02, PappuPhD,Buch05,Iesl16}.  
If the verification set-up is not tamper-resistant, such EAPs are 
vulnerable to emulation attacks, in which the adversary has stolen the PIN and knows the challenge-response 
properties of the PUK. Hence, his task is to intercept each challenge during the  
verification stage, measure it, and send to the verifier the expected response \cite{Goorden14}.  
This is a highly realistic scenario \cite{remark1}, and if the adversary can access  the classical 
challenges during the verification, nothing prevents him from determining them completely, 
without introducing any errors. 

The robustness of PUK-based optical EAPs against emulation attacks can be increased considerably 
if the PUK is challenged by quantum states of light. When the quantum state is chosen at random 
from a set of non-orthogonal  states, the laws of quantum physics prevent an adversary to 
discriminate unambiguously 
between different states, while information gain can be obtained at the cost of  disturbing the state of the 
probe \cite{book}. Hence, if the EAP is designed judiciously so that  the response of the PUK depends on 
the quantum state of the probe, 
then it is highly unlikely for an adversary to estimate correctly all of the challenges 
used during the verification of a PUK without introducing any errors. 
In this spirit, Goorden {\em et al} have proposed 
and implemented an authentication protocol, in which the challenges pertain to attenuated laser pulses 
with shaped wavefronts \cite{Goorden14}. The mean number of photons per mode in the 
challenge is less than 1, which prevents an adversary from characterizing fully the 
wavefront of the challenge. 

Recently, we proposed an alternative  PUK-based optical EAP, which relies on standard wavefront-shaping 
and homodyne-detection techniques \cite{NikDiaSciRep17}. In the framework of a tamper-resistant 
verification set-up, it was shown that the protocol offers  collision resistance, and it is robust against PUK cloning. The present work focuses on the verification stage of the protocol, and in particular on 
the authentication of the PUK to the system. Our main task is to investigate  the robustness of the 
protocol against an emulation attack, in which a cheater has obtained access to the database without 
being noticed \cite{remark1}, and moreover he has been able to bypass the actual verification 
set-up, thereby directing each probe to his detection device for analysis.

%%%%%%%%%%%%%%%%%%%%%%%%%
%             SECTION  II
%%%%%%%%%%%%%%%%%%%%%%%%%

\section{Authentication scheme}
\label{sec2}

The verification set-up has been discussed in detail elsewhere \cite{NikDiaSciRep17}, and for the sake 
of completeness, its main components are  outlined in Fig. \ref{fig1}.
 It consists of three major parts namely, the laser source, the interrogation chamber, and the 
 homodyne-detection chamber. 
Light from the laser source is split into two parts: a weak probe that is directed to the interrogation 
chamber by means of a single-mode fiber (SMF), and a strong local oscillator (LO), which serves as a 
reference in the homodyne detection (HD). 
The interrogation chamber is basically a  standard wavefront-shaping set-up, in which the light from SMF A 
is collimated, and its spatial phase pattern is modulated by means of a phase-only spatial 
light-modulator (SLM) \cite{Vellekoop07,Vellekoop08,Mosk12,Huisman14,Huisman15,Poppoff10,Poppoff11,Defienne14,Huisman14b,Wolterink16}. 
The shaped wavefront is then focused on the PUK, and the phase mask of the SLM 
is optimized so that the multiple-scattered light is mainly directed to a prescribed transversal spatial 
mode of the output plane (target mode), where it is coupled to SMF B \cite{Vellekoop07,Vellekoop08,Mosk12,Huisman14,Huisman15,Poppoff10,Poppoff11,Defienne14,Huisman14b,Wolterink16}.  The collected scattered light is transferred to the HD set-up, where its quadratures are  analysed. 

Standard wavefront shaping set-ups allow for the control of hundreds and even thousands of spatial 
modes,  and the intensity of the scattered light is enhanced relative to the unoptimized case by a factor 
${\cal E}\gg 1$. Due to noise and imperfections, the enhancement factor that can be achieved in practise 
is below the ideally expected  value of $\pi{{\cal N}}/4 \simeq 0.78 {\cal N}$ 
\cite{Vellekoop08,Poppoff11,Yilmaz13,Anderson14}, 
where ${\cal N}$ is the number of modes that can be controlled in the wavefront-shaping set-up. 
The reported ratios ${\cal E}/{\cal N}$  in related experiments range from about $0.2$ to about 
$0.6$ \cite{Yilmaz13}. 

\subsection{Challenge-response pairs}
\label{sec2a}

The enrolment of a PUK takes  place only once, and aims at the generation of a finite set of CRPs, 
which is stored in a server, and will be used for the authentication of the PUK. 
The enrolment is performed by the manufacturer well before the PUK is given to a user, and we naturally assume that the enroller has all the time and the resources needed so 
that the accuracy in the estimation of the response of the PUK to each one of the challenges, is considerably higher than 
the accuracy in the verification stage \cite{NikDiaSciRep17}.  Throughout this work we consider identical set-ups for the enrolment and the verification stages, which operate in the diffusive limit \cite{Goodman1,book2}, and all of their specifications (i.e., transmission losses, imperfections, etc) are publicly known. 

Our scheme accepts  various types of challenges, pertaining to the probe state, the target mode and the 
phase mask of the SLM.   In this work, we consider a rather simple challenge, by assuming that the  target 
mode (i.e., the position of SMF B at the output plane) is fixed and publicly known. Hence, a challenge has two parts and it  is of the form 
\bea
C_k =\left \{k, {\bm \Phi}({\cal K}) ~:~k\in \Int_N\right \}, 
\label{challenge:eq}
\eea
with the integer $k$ chosen at random from a uniform distribution over $\Int_N \equiv \{0,1,2,\ldots,N-1\}$, while the integer $N\gg 1$ is a publicly known constant. We consider coherent probe states with the same, publicly known mean number of photons $\mu_P$, and random phase $ \varphi_k$ i.e., 
\bea
\ket{\alpha_k}:=\ket{\sqrt{\mu_P}e^{{\rm i}\varphi_k}},\,\textrm{with}\,\, \varphi_k:=\frac{2\pi k}{N}.
\label{alpha:eq}
\eea  
Hence, the integer $k$ identifies uniquely the probe state. 
Finally,  ${\bm \Phi}$ denotes the optimal phase-mask for the SLM, which maximizes the intensity of the scattered light in SMF B, and depends on the scattering matrix of the PUK ${\cal K}$, as well as various  parameters of the fixed set-up, which are assumed to be publicly known and are not shown in Eq. (\ref{challenge:eq}). It is worth emphasizing here that for the optimization process used in our 
simulations \cite{NikDiaSciRep17},  the optimal phase mask of the SLM does not depend on the 
phase of the probe \cite{remark4}. 
In fact, it has been demonstrated both theoretically and experimentally  
that an optimal phase mask works equally well, regardless the quantum state of the incoming light  \cite{Defienne14,Huisman14b,Wolterink16}. 

The response of the PUK to a given challenge $C_k$, pertains to the electric field 
$\hat{E}_k({\cal K}) := \hat{X}_k({\cal K}) + {\rm i}\hat{Y}_k({\cal K})$ in  SMF B (see table \ref{tab1}). 
For later convenience, let us introduce here the generalized $\theta-$quadrature $\hat{Q}_k(\theta)$ of the field in SMF B, with $\hat{Q}_k(0) = \hat{X}_k$ and $\hat{Q}_k(\pi/2) = \hat{Y}_k$. 
It has been shown that in the diffusive limit, the field in SMF B is also in a coherent state, say 
$\ket{\beta_{k}}$, with the expectation value of $\theta-$quadrature given by (see also appendix \ref{app1}) \cite{NikDiaSciRep17} 
\begin{subequations}
\label{Q_theta:eq}
\bea
\aver{\hat{Q}_k(\theta)} := \sqrt{2 \mu_R} \cos (\psi_k-\theta),  
\label{Q_theta1:eq}
\eea
with  
\bea
\label{mu_R:eq}
&&\mu_R ={\cal E} |{\cal F}|^2\mu_P, \\
&&\psi_k:= \arg({\cal F} )+\varphi_k.
\label{psi_k:eq}
\eea
\end{subequations}
The factor ${\cal F}$  is a complex number that depends on the PUK ${\cal K}$, the phase-mask of the SLM and various other publicly-known parameters of  the set-up, while $|{\cal F}|^2<1/{\cal N}$ \cite{NikDiaSciRep17}. As mentioned above, the enhancement factor ${\cal E}$ 
depends on the details of set-up, and in practise it does not exceed the ideal value of $\pi{\cal N}/4$. 
Hence,  we expect $\mu_R < \pi \mu_P /4\approx 0.78\mu_P$. 
For a fixed verification set-up and a given PUK, $\arg({\cal F} )$ is a global phase, which shifts the phase of the probe $\varphi_k$, and thus it is not expected to affect the following security analysis (see related discussion in Sec. \ref{sec3c}).  In the  attack to be considered below, the adversary is assumed to know the CRPs for  the PUK \cite{remark1}, while ${\cal F}$ is 
publicly known. 

\begin{table}
\centering
\caption{Illustration of a set of CRPs used for authentication of a PUK. The angles $\theta = 0$ and $\pi/2$ refer 
to the quadratures $\hat{X}_k$ and $\hat{Y}_k$  of the response field $\hat{E}_k$, respectively.}
\label{tab1}
%\begin{ruledtabular}
\begin{tabular}{c|c|c|c}
\hline
\hline
\multicolumn{4}{c}{PIN}  \\
\hline
 \multicolumn{2}{c|}{Challenge}    & \multicolumn{2}{c}{Response}             \\ 
\hline
   $k$ & Phase mask  & $\theta=0$  & $\theta = \pi/2$ \\ \hline
 0 &   & $\aver{\hat{X}_0}$  & $\aver{\hat{Y}_0}$\\ 
 1  &   & $\aver{\hat{X}_1}$ & $\aver{\hat{Y}_1}$\\ 
 2 & ${\bm \Phi}$  & $\aver{\hat{X}_2}$  & $\aver{\hat{Y}_2}$\\
 $\vdots$ &   & $\vdots$  & $\vdots$\\ 
 $N-1$ &   & $\aver{\hat{X}_{N-1}}$  & $\aver{\hat{Y}_{N-1}}$\\
 \hline
 \hline
\end{tabular}
%\end{ruledtabular}
\end{table}
%%%%%%%%%%%%%

For a fixed set-up and given PUK, a CRP is uniquely identified by the choice of the probe 
i.e., by the choice of the integer 
$k\in\Int_N$, because throughout this work, all of the parameters of the set-up, including 
$\mu_P$ and  the position of SMF B, are assumed to be publicly known.  
As will be explained below, even if an adversary has obtained access to the server where the database 
of CRPs is stored, the response of the PUK  to a randomly chosen probe state $\ket{\alpha_k}$ cannot 
be estimated without knowledge of the integer $k$. 

To suppress notation, the dependence of ${\bm \Phi}$ and $\hat{E}_k$ on the PUK  ${\cal K}$, is not explicitly shown in the following sections. In the emulation attack under consideration the attacker bypasses the interrogation chamber, and thus 
the PUK does not enter the related security analysis.  
Before we proceed with the attack, however, it will be helpful to discuss briefly the verification of a PUK, in the ideal scenario of tamper-resistant verification set-up and secure server. 

\subsection{Verification in the absence of cheating}
\label{sec2b}

By contrast to the enrolment stage, verification takes place each time a PUK has to be authenticated by a legitimate verifier, who has access to the server (where the database of CRPs is located) over a secure and authenticated classical channel.  The first action of the verifier is to send the user's PIN to the server, over the classical channel. If the PIN is not valid the verification is aborted. Otherwise, the server returns a number of challenges, say $\{C_1, C_2,\ldots,C_M\}$ [see Eq. (\ref{challenge:eq})], chosen at random and independently from the  registered finite set of CRPs for the PUK to be authenticated. In addition, the server sends to the verifier a sequence of random and independently chosen angles $\{\theta_1, \theta_2,\ldots,\theta_M\}$, with $\theta_j$ uniformly distributed over 
$\{0,\pi/2\}$ \cite{remark2,remark6}. 

The verifier queries the PUK with each one of the $M$ challenges sequentially, and records the corresponding responses. Given that the 
challenges are chosen at random and independently, it is sufficient to consider one of the challenges, say the $j$th one.  
The phase mask of the SLM is set to ${\bm \Phi}$, 
and the phase of the probe is set to $\varphi_{k_j} = 2\pi k_j/N$. 
The verifier measures the quadrature amplitude of the electric field of the scattered light in SMF B, by means of  HD where the LO serves as the required reference \cite{book3}. 
More precisely, for the $j$th query,  the LO phase is set to 
$\theta_j\in\{0,\pi/2\}$, and the verifier measures the 
quadrature $\hat{Q}_{k_j}(\theta_j)$. 
Assuming that the LO field is much stronger than the total scattered field,  the outcome of such a measurement is a real random number 
$q_j$ which, to a good accuracy, follows a normal (Gaussian) distribution ${\mathscr N}(\bar{q}_{k_j}(\theta_j),\sigma^2)$, with standard deviation $\sigma \simeq 1/\sqrt{2\eta}$, where  $\eta\leq 1$ is  the detection efficiency \cite{Raymer95}.  
In other words,  the measurement of the quadrature $\hat{Q}_{k_j}(\theta_j)$ is equivalent to sampling from the Gaussian 
distribution ${\mathscr N}(\bar{q}_{k_j}(\theta_j),\sigma^2)$. The mean 
$\bar{q}_{k_j}(\theta_j)$ also depends on the scattering matrix of the PUK,  
but for the reasons discussed above, this dependence is not explicitly shown. The  sequence of  random outcomes  $\{q_1, q_2,\ldots,q_M\}$,  are sent to the server over the secure and authenticated classical channel. 

Aiming at practical verification tests, and in order to circumvent difficulties arising from statistical 
deviations in finite sampling, we proposed a verification process that employs two-bin coarse-graining 
of the outcomes from the $M$ independent HDs. 
More precisely, in the $j$th sample with challenge $C_j$ and measured 
quadrature $\hat{Q}_{k_j}(\theta_j)$,  the server checks if the outcome $q_j$ falls in the interval (bin) 
\bea
{\mathbb B}_{k_j}(\theta_j) = 
\left [
\aver{\hat{Q}_{k_j}(\theta_j)}-\frac{\Delta}2,
~\aver{\hat{Q}_{k_j}(\theta_j)}+\frac{\Delta}2 
\right ], 
\label{bin:eq}
\eea
which is centred at the expected response recorded in the database of CRPs, and has size $2\sigma \lesssim \Delta<4\sigma$.

\begin{figure*}
\includegraphics[scale=0.6]{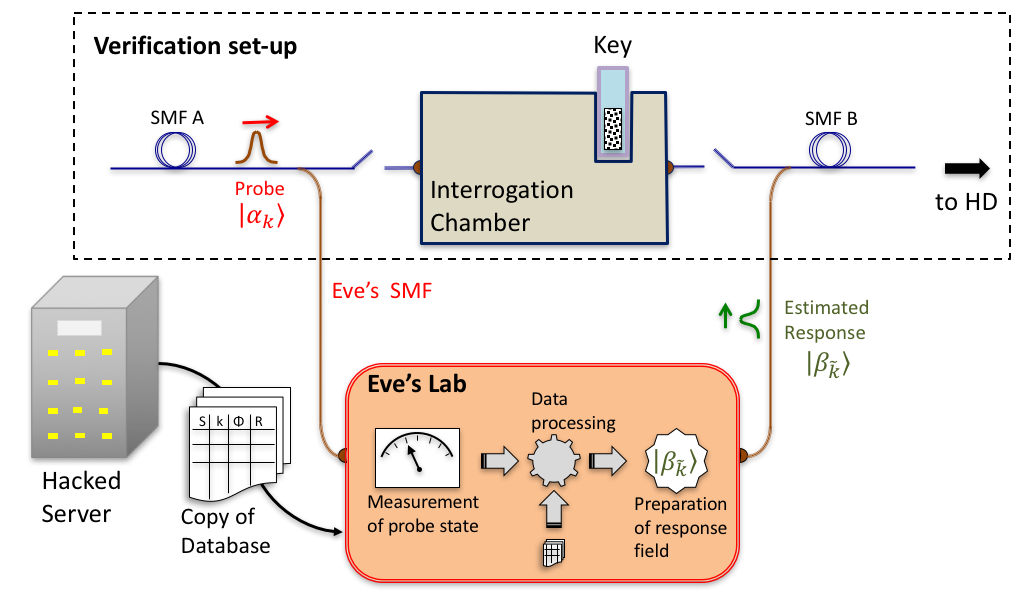}
\caption{(Color online) A schematic representation of the emulation attack considered throughout this work. An adversary who knows the set of all possible CRPs pertinent to a PUK, but he does not possess the actual PUK, bypasses the interrogation chamber, thereby directing each probe in his ideal lab \cite{remark3}. 
The probe is measured, and a response state, which is consistent with the outcome of the measurement and 
the list of possible CRPs,  is sent to the HD set-up of the verifier (not shown here). 
}
\label{fig2}
\end{figure*}

In the ideal scenario, where the true PUK is authenticated in a tamper-resistant verification set-up 
and no cheating takes place,   the verifier samples from a Gaussian distribution centred at 
$\bar{q}_{k_j}(\theta) = \aver{\hat{Q}_{k_j}(\theta_j)}$, which coincides 
with center of the bin $ {\mathbb B}_{k_j}(\theta_j) $. 
Hence, the  probability for the outcome  $q_{k_j}$ to fall in the bin, given that 
the PUK is interrogated by challenge $C_{j}$ and the $\theta_j-$quadrature  is measured, 
is given by (see appendix \ref{app2}) 
\bea
P^{(0)}({\rm in}|k_j,\theta_j) =  {\rm Erf}\left (\frac{\bar{\Delta}}{2\sqrt{2}}\right ).  
\label{Pexp:eq}
\eea
The key point is that this probability is independent of $k_j$ and $\theta_j$, and depends only on the ratio $\bar{\Delta}:=\Delta/\sigma$. 

Counting the total number of queries (samples) $M_{\rm in}$ that resulted in an outcome within the expected bin, irrespective of the chosen challenge or the measured quadrature, the server  obtains the relative frequency  $f_{\rm in} := M_{\rm in}/M$,  which is an estimate of the average probability 
\bea
P_{\rm in}^{(0)} &:=& \sum_{k=0}^{N-1} ~ \sum_{\theta\in{\{0,\pi/2\}}} p(k) p(\theta) P^{(0)} ({\rm in}|k,\theta) 
\nonumber \\
&= &
\frac{1}{2N}\ \sum_{k=0}^{N-1}~\sum_{\theta\in\{0,\pi/2\}} P^{(0)} ({\rm in}|k,\theta). 
\label{av_prob2:eq}
\eea
Here, we have taken into account that $k$ and $\theta$ are chosen at random and independently from uniform distributions over $\Int_N$ and $\{0,\pi/2\}$, respectively. 
Substituting Eq. (\ref{Pexp:eq}) in Eq. (\ref{av_prob2:eq}), one readily obtains 
\bea
P_{\rm in}^{(0)} = {\rm Erf}\left (\frac{\bar{\Delta}}{2\sqrt{2}}\right ).
\label{av_prob3:eq}
\eea 

The acceptance or rejection of the PUK is decided upon the deviation of the estimated probability $f_{\rm in}$ from the theoretically expected one, which in the absence of any cheating is given by  $P_{\rm in}^{(0)}$. 
It has been shown that a sample of size at least as large as 
\bea
M_{\rm th} := \frac{3\ln(2\zeta^{-1})}{\varepsilon^{2}}, 
\label{Mth:eq}
\eea
suffices for the estimate $f_{\rm in}$ to satisfy 
\bea
{\rm Pr}(|f_{\rm in} - P_{\rm in}^{(0)}|\geq \varepsilon)<\zeta, 
\eea
where $\zeta\ll 1$ is the uncertainty, and $\varepsilon\ll P_{\rm in }^{(0)}$ is the absolute 
error \cite{NikDiaSciRep17}. 
For $M>M_{\rm th}$ one can be $100(1-\zeta)\%$ confident that in the absence of 
any cheating,  the estimate will lie within an interval of size $2\varepsilon$ around the theoretically 
expected probability i.e., $|f_{\rm in} - P_{\rm in}^{(0)}|< \varepsilon$.  
So, a PUK is accepted if  $|f_{\rm in} - P_{\rm in}^{(0)}|< \varepsilon$, and is rejected otherwise.  

According to Eq. (\ref{Mth:eq}), samples of size $M \simeq 2.3\times 10^{7}$  suffice for verification tests of error $\varepsilon\simeq 10^{-3}$ and confidence 99.9\%. Assuming that the separation distances between the different  components of the verification set-up are of the order of meters or tens meters, and for HD bandwidth $\sim 10 - 100$ MHz (depending on the implementation), the total verification time is expected to be a  few seconds \cite{NikDiaSciRep17}. 

The interested reader may refer to Ref. \cite{NikDiaSciRep17}, for a detailed discussion on the different stages of the protocol, a full description of its implementation, as well as for a thorough analysis on its security in the case of  tamper-resistant verification set-up and secure server. 
Both of these assumptions are relaxed in the emulation attack discussed in the following section. 
In this case, the cheater's intervention is expected to introduce errors in the responses of the PUK to the various challenges, thereby enforcing the verifier to sample from photocount distributions, which are shifted relative to the expected bins. The protocol is secure only if the aforementioned 
coarse-grained HD can detect these shifts, which is not true for any combination of $\mu_P$ and 
$N$.  For the sake of convenience,  the main parameters entering the following security analysis are 
summarized in table \ref{tab2}. 

\begin{table}
\centering
\caption{List of the main parameters in the present security analysis. 
The PIN as well as the scattering matrix of the PUK and/or the challenge-response pairs are assumed to have  leaked to the adversary (see Sec. \ref{sec1} and 
remark \cite{remark1}). }
\label{tab2}
%\begin{ruledtabular}
\begin{tabular}{c|c|c}
\hline
\hline
Parameter  & Symbol  & Property   \\ \hline
Mean number of photons in probe &$ \mu_P$   & Public \\ 
Phase of probe & $\varphi_k$    & Private \\ 
Total number of  phases  & $N$ & Public \\
 Mean number of photons in response  & $\mu_R$ & Public  \\
 PIN  & - & Leaked  \\
 Position of SMF B  & - & Public  \\
 \hline
PUK & ${\cal K}$ & Private\\
Scattering matrix of PUK & -  & Leaked\\
 \hline
 Number of controlled modes & ${\cal N}$ & Public\\
Phase mask of SLM &  ${\bm \Phi}$ & Leaked\\
Enhancement factor & ${\cal E}$    & Public \\ 
Detection efficiency & $\eta$ & Public\\ 
 -  & ${\cal F}$ & Public  \\
Other specifications of set-up & - & Public  \\
 \hline
Bin size & $\Delta$ & Public \\
Error in verification tests & $\varepsilon$ & Public\\
Confidence in verification tests & $\zeta$ & Public\\
Challenge-response pairs & CRPs & Leaked\\ 
 \hline
 \hline
\end{tabular}
%\end{ruledtabular}
\end{table}
%%%%%%%%%%%%%

%%%%%%%%%%%%%%%%%%%%%%%%%
%             SECTION  III
%%%%%%%%%%%%%%%%%%%%%%%%%

\section{Emulation attack and security}  
\label{sec3}

Consider the cheating strategy of Fig. \ref{fig2}, in which the adversary aims at the impersonation of the  
legitimate user, without having the true PUK. 
The adversary does not have access either to the laser source or to the HD set-up of the verifier, but he has hacked the server, and has obtained the PIN of the user he intends to impersonate, together with a copy of the CRPs of his PUK \cite{remark1}. Moreover, he has installed his own perfect line, thereby bypassing the verification set-up, and directing each probe to his lab. Clearly, there is no need for the adversary to have the user's PUK, 
because he has a copy of the corresponding list of CRPs \cite{remark3}.
 If the cheater can estimate reliably the probe state, he can prepare the right response by looking up the set of all possible CRPs, and send to  the HD set-up of the verifier a quantum state that is consistent with the expected response of the true PUK. This is possible for classical challenges, because nothing prevents the cheater from estimating precisely all of the characteristics of the probe, without introducing any errors. Hence, any EAP where the PUK is challenged by classical signals \cite{Horstmeyer15,Pappu02, PappuPhD,Buch05,Iesl16} is susceptible to such an  emulation attack. The situation is fundamentally 
 different in our protocol, because the PUK is interrogated by random non-orthogonal quantum states. 
According to quantum mechanics, non-orthogonal quantum states cannot be reliably distinguished \cite{book}, and thus the adversary's intervention is expected to introduce errors in the 
coarse-grained HD.

%=====================
%             Sub-Section  IIIa
%=====================

\subsection{Error probability}
\label{sec3a}

In our scheme, the probe state is a pure coherent state $\ket{\alpha_k}$, with  random phase. 
From the adversary's point of view, before any measurements,  the probe's state is a mixture of the form 
\bea
\rho
= \frac{1}{N}
\sum_{k=0}^{N}\ket{\alpha_k}\bra{\alpha_k}.
\label{rho:eq1}
\eea

\begin{figure*}
\includegraphics[scale=0.5]{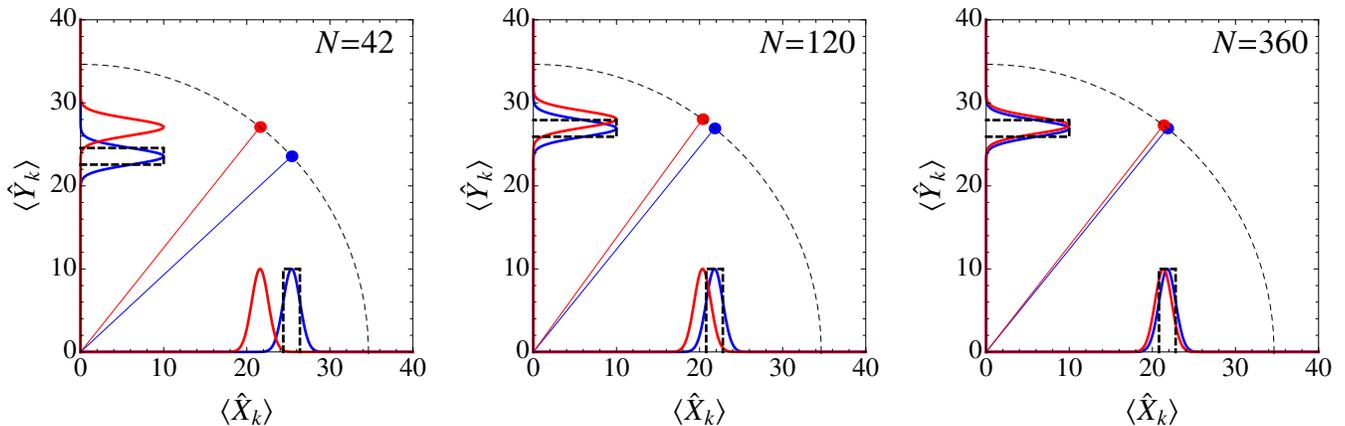}
\caption{(Color online) Two neighbouring coherent states  in phase-space representation (blue and red disks), with the same mean number of photons $\mu = 600$, and phases that differ by $2\pi/N$,   for three different values of $N$. The radius of the disks is equal to the shot noise i.e., $1/\sqrt{2}$. The Gaussians show the corresponding photocount distributions for each quadrature, and their standard deviation is $\sigma = 1/\sqrt{2\eta}$, for $\eta= 0.5$. The dashed box represents an interval of size $\Delta = 2\sigma$. 
}
\label{fig3}
\end{figure*}

The average information that an adversary can extract from the state  (\ref{rho:eq1})  is limited by the 
Holevo bound $\chi$, which is given by  the von Neumann entropy 
$S(\rho)\equiv {\rm Tr}[\rho\log_2(\rho)]$ i.e., 
\bea
\chi = S(\rho). 
\label{chi:eq}
\eea
For given $\mu_P$ and $N$, the entropy $S(\rho)$ can be calculated numerically. 
A simple analytic expression  can be obtained in the continuous limit 
$N\to \infty$, where 
$S(\rho)$ approximates the entropy of the Poisson distribution, which for for $\mu_{P}\gtrsim 10$ is given by 
\bea
 {\mathscr H}_{\infty}(\mu_{P})  \simeq  \log_2(\sqrt{2\pi e\mu_{P}}). 
\eea

The overlap of neighbouring probe states is determined by the  number of different  phases $N$, and the mean number of photons $\mu_P$. The larger $N$ becomes for a given $\mu_P$, the larger the overlap becomes (see Fig. \ref{fig3}). 
%The continuum limit pertains to $N\gtrsim 2\pi\sqrt{2 \mu_{P}}$. 
 
Our scheme involves $M\gg 1$ equivalent samples (queries), and for each one of them the positive integer $k$ is chosen at random and independently from $\Int_N$.   In the emulation attack under consideration, the adversary has a copy of all the relevant CRPs. Hence, for each one of the samples, he intercepts the probe state and measures it, so that to obtain an estimate of the integer $k$, say $\widetilde{k}$. By looking up the set of CRPs, he finds the expected response $\{\aver{\hat{X}_{\widetilde k}},\aver{\hat{Y}_{\widetilde k}}\}$ that corresponds to $\widetilde{k}$. Subsequently, he prepares and sends to the HD (through SMF B ) the coherent state $\ket{\beta_{\widetilde k}}$ which satisfies $\bra{\beta_{\widetilde k}} \hat{E}\ket{\beta_{\widetilde k}} =\aver{\hat{X}_{\widetilde k}}+{\rm i}\aver{\hat{Y}_{\widetilde k}}$. 

It may happen that in a single sample the adversary will obtain the right value of $k$. 
In this case, the centres of the Gaussian photocount distributions for the two quadratures coincide with the centres of the corresponding bins i.e., $\aver{\hat{X}_{k}}$ and $\aver{\hat{Y}_k}$.  However, if $\mu_P$  and $N$ are such that there is significant overlap between different possible probe states, it is impossible for the adversary to distinguish them reliably, and thus to  obtain the right value of $k$ in all of the $M$ queries. In other words, it is inevitable for the adversary to make wrong guesses about the state of the probe in some queries, and thus 
send to the HD set-up a response state $\ket{\beta_{\widetilde k}}$ for which 
the centres of the photocount distributions are shifted relative to the expected bins i.e., 
$\aver{\hat{X}_{\widetilde k}} \neq  \aver{\hat{X}_{k}}$ and $\aver{\hat{Y}_{\widetilde k}} \neq \aver{\hat{X}_k}$. In these cases, the probability for the outcome of the HD to fall in the bin is smaller than the probability of Eq. (\ref{Pexp:eq}). 

It is sufficient for our purposes to consider the  probability $p_{\rm err}$ 
for the adversary to infer a wrong value of $k$, based on the outcome of his measurement on probe state $\ket{\alpha_k}$, and perhaps some additional post processing.  According to Fano's inequality, 
when $k$ is uniformly distributed on $\Int_N$ 
the probability of error satisfies  \cite{book}
\bea
{\mathscr H}(p_{\rm err} )+p_{\rm err} \log_2(N-1)
 &\geq& \log_2(N)-\chi,   
 \label{fano1:eq}
\eea
with the Holevo bound given by Eq. (\ref{chi:eq}). 
Solving this inequality for a given state $\rho$, one obtains a lower bound  on the error probability, say $p_{\rm err}^{({\rm low})}$.

%=====================
%             Sub-Section  IIIb
%=====================

\subsection{Security condition}
\label{sec3b}

As discussed in Sec. \ref{sec2b}, in the EAP under consideration 
acceptance or rejection of a PUK is decided upon the estimated 
average probability for an outcome to fall in a bin. Given that the cheater's intervention will inevitably 
introduce errors, we need to find how these errors affect the estimated probability in the verification stage. 

Consider a single query, where the verifier sends $\ket{\alpha_k}$,  the cheater infers 
$\widetilde{k}$ and sends 
the coherent state $\ket{\beta_{\widetilde{k}}}$ to the HD set-up of the verifier. 
As discussed above, the wrong guess of the cheater will enforce the verifier to sample from a 
Gaussian photocount distribution that is shifted relative to the expected bin. More precisely, 
the conditional probability for the outcome of the HD to fall in the expected bin ${\mathbb B}_k(\theta)$, 
given that the verifier has chosen $k$, the adversary has inferred $\widetilde{k}$, and the verifier measures the $\theta$ quadrature of the field is 
given by  (see appendix \ref{app2})
\bea
P({\rm in}|k,\widetilde{k},\theta) &=& 
\frac{1}{2}\left \{ {\rm Erf} \left [
\frac{2 S(k,\widetilde{k},\theta) +\Delta}{2\sqrt{2}\sigma}
\right ]\right.
\nonumber \\
&&
-
\left. {\rm Erf} \left [
\frac{2S(k,\widetilde{k},\theta)-\Delta}{2\sqrt{2}\sigma}
\right ]\right \},
\label{Pexp_cheating:eq}
\eea
where 
\bea
S(k,\widetilde{k},\theta):=\aver{\hat{Q}_{\widetilde{k}}(\theta)}-\aver{\hat{Q}_k(\theta)}, 
\label{shift:def}
\eea
and 
$\aver{\hat{Q}_{\widetilde{k}}(\theta)}$ is also given by Eq. (\ref{Q_theta:eq}), for index  
$\widetilde{k}$ instead of $k$. Recall here that $\aver{\hat{Q}_k(\theta)}$ is the centre of the bin when 
the probe state $\ket{\alpha_k}$ has been used and the $\theta-$quadrature  is measured 
(see Eq. \ref{bin:eq}). When the adversary infers the correct value of $k$ i.e., for $\widetilde{k} = k$,  
then $\aver{\hat{Q}_k(\theta)}$ coincides with the mean of the Gaussian  photocount distribution the 
verifier samples from. When the adversary fails to infer the correct value of $k$, i.e., 
for $\widetilde{k} \neq k$, the photocount distribution is centred at $\hat{Q}_{\widetilde{k}}(\theta)$. 
Hence, $S(k,\widetilde{k},\theta)$ is the shift of the photocount distribution relative to the bin, which is zero only in the case of $\widetilde{k} = k$, and non-zero otherwise. In the former case, Eq. (\ref{Pexp_cheating:eq}) reduces to Eq. (\ref{Pexp:eq}), which shows that the cheater's intervention has not affected the photocount statistics expected by the server in the absence of cheating. 

Given that the server asks the verifier to sample from both quadratures at random and independently, after $M$ queries 
the verifier will obtain an estimate of the average probability 
\bea
P_{\rm in} = \sum_{k=0}^{N-1} \sum_{\widetilde{k}=0}^{N-1} \sum_{\theta} P({\rm in},k,\widetilde{k},\theta).  
\eea
The arguments of Sec. \ref{sec2b} are also valid in the case of cheating. That is,  
as long as $M>M_{\rm th}$ and $\varepsilon\ll P_{\rm in}$, with high probability 
the estimate of the server will lie within an interval of size $2\varepsilon$ around the theoretically expected probability, which in the presence of cheating is $P_{\rm in} $ instead of $P_{\rm in}^{(0)}$ 
 i.e., we have 
\bea
|f_{\rm in} - P_{\rm in}|<\varepsilon,
\label{stdev:cheat}
\eea 
with high probability. 

The probability $P_{\rm in} $ is different from the corresponding probability in the absence of cheating $P_{\rm in}^{(0)}$, and can be bounded from above as follows  (see appendix \ref{app3})
\bea
P_{\rm in}  \leq   (1-p_{\rm err}) P_{\rm in}^{(0)}+p_{\rm err} 
  \max_{k,\widetilde{k}}  \{ P({\rm in}|k,\widetilde{k})\}_{k\neq\widetilde{k}}, 
 \label{Pin_cheat:ineq}
\eea
where $P_{\rm in}^{(0)}$ is given by Eq. (\ref{av_prob3:eq}) and 
\[
P({\rm in}|k, \widetilde{k}) := \sum_{\theta\in\{0,\pi/2\}} p(\theta) P({\rm in}|k,\widetilde{k},\theta).
\]
Recalling that both quadratures are equally probable, and using 
Eqs. (\ref{Pexp_cheating:eq}), (\ref{shift:def}) we have 
\begin{widetext}
\bea
P({\rm in}|k, \widetilde{k})\} 
&=& 
\frac{1}4 \sum_{\theta\in\{0,\pi/2\}} \left \{ { \rm Erf} \left [
\frac{2 (\aver{\hat{Q}_{\widetilde{k}}(\theta)}-\aver{\hat{Q}_k(\theta)})+\Delta}{2\sqrt{2}\sigma}
\right ]
-
{\rm Erf} \left [
\frac{2(\aver{\hat{Q}_{\widetilde{k}}(\theta)}-\aver{\hat{Q}_k(\theta)})-\Delta}{2\sqrt{2}\sigma}
\right ]\right \}.
 \label{theta_av_P:eq}
\eea
\end{widetext}
where $\aver{\hat{Q}_{\widetilde{k}}(\theta)}$ and $\aver{\hat{Q}_k(\theta)}$ are given by 
Eq.  (\ref{Q_theta:eq}) for $\widetilde{k}$ and $k$, respectively. 
Moreover, using Eqs. (\ref{av_prob3:eq}) and (\ref{Pexp_cheating:eq}), one can readily confirm that 
$\max_{k,\tilde{k}}  \{ P({\rm in}|k,\tilde{k})\}_{k\neq\tilde{k}}\leq P_{\rm in}^{(0)}$, 
and thus 
\bea
P_{\rm in}^{(0)}- P_{\rm in}\geq p_{\rm err} \left ( P_{\rm in}^{(0)}-
  \max_{k,\tilde{k}}
  \{P({\rm in}|k,\tilde{k})\}_{k\neq\tilde{k}} \right )\geq 0,
  \nonumber \\
  \label{prob-diff:eq}
\eea
where equality holds only in the absence of cheating.  
 
As described in Sec. \ref{sec2b},  a PUK is accepted only if the estimate $f_{\rm in}$  
lies within an interval of size $2\varepsilon\ll 1$ around $P_{\rm in}^{(0)}$, and not around $P_{\rm in}$. 
So, the cheating will be detected if  
\bea
P_{\rm in}^{(0)}-P_{\rm in}>2\varepsilon. 
\label{cheating_detection:eq}
\eea
Indeed, in view of inequality (\ref{stdev:cheat}), inequality (\ref{cheating_detection:eq}) implies 
$P_{\rm in}^{(0)}-f_{\rm in}>\varepsilon$, with high probability. That is, the     detected deviations from 
$P_{\rm in}^{(0)}$ exceed the statistical deviations, and can be attributed only to some sort of cheating. 
Of course, one cannot infer the precise type of the cheating, but this is not of importance for the security of the 
protocol. 

From inequalities (\ref{prob-diff:eq}) and (\ref{cheating_detection:eq}), we obtain that the condition    
\bea
p_{\rm err}^{({\rm low})}\left (
P_{\rm in}^{(0)} - \max_{k,\tilde{k}}
  \{P({\rm in}|k,\tilde{k})\}_{k\neq\tilde{k}} \right  )>2\varepsilon, 
  \label{security_cond}
\eea
suffices to ensure the security of the protocol, where $p_{\rm err}^{({\rm low})}$ 
is a lower bound on the error probability $p_{\rm err}$ obtained through inequality (\ref{fano1:eq}). 

Condition (\ref{security_cond}) is the main result of this work. The left-hand side (l.h.s) of the inequality  is 
a function of $\mu_P$, $N$, $\mu_R$, $\eta$, $\bar{\Delta}$, and for later convenience it will be 
denoted by $D$. For a fixed verification set-up, with known parameters, condition 
(\ref{security_cond}) allows one to estimate combinations of $\mu_P$, $N$ and $\varepsilon$, so that 
the protocol is secure against the emulation attack under consideration, 
in the sense that  the cheating will be detected, and the adversary will fail to pass himself  as the legitimate user.

\begin{figure*}
\centerline{\includegraphics[scale=0.45]{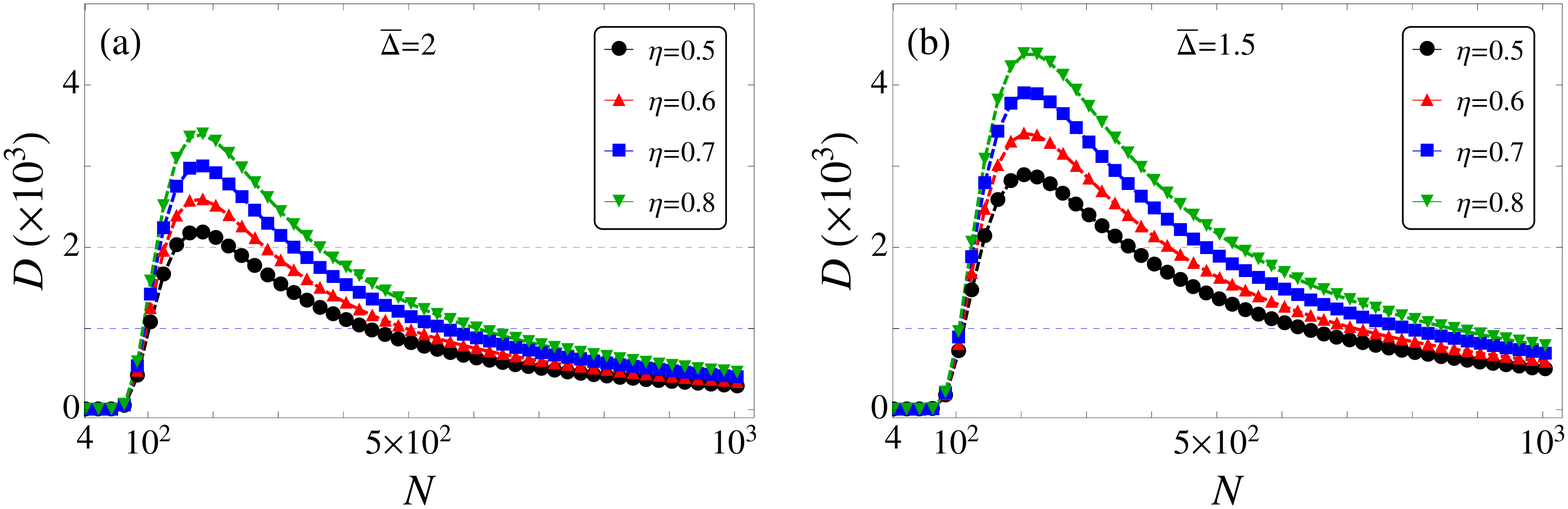}}
\centerline{\includegraphics[scale=0.45]{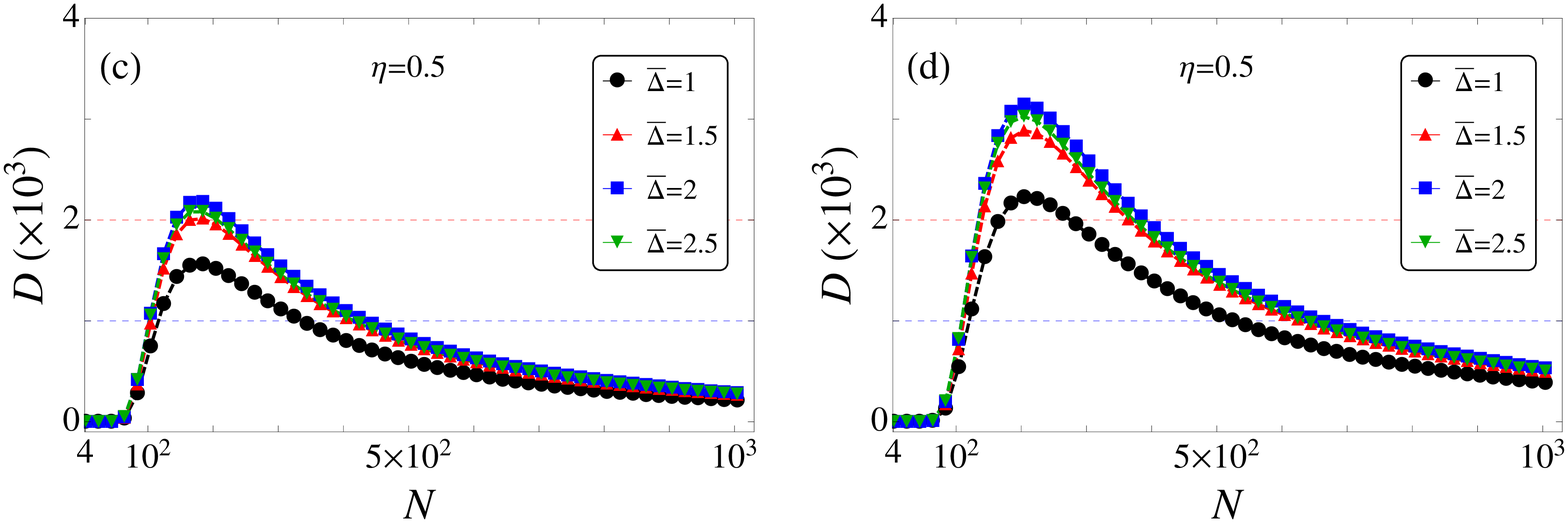}}
\caption{(Color online) (a,b) The l.h.s of inequality (\ref{security_cond}) is plotted as a function of the number of phases $N$, for various values of $\eta$ and two different values of $\bar{\Delta}$.  The horizontal blue and red dashed lines correspond to the right-hand side of (\ref{security_cond}) for $\varepsilon=5\times 10^{-4}$ and $\varepsilon = 10^{-3}$, respectively.  The protocol is secure for values $N$, for which $D$ exceeds $2\varepsilon$.   
(c,d) Same as in plots (a,b), but for various values of $\bar{\Delta}$, and fixed $\eta=0.5$. 
Other parameters: (a,c) $\mu_P = 600$, $\mu_R=0.2\mu_P$; (b,d) $\mu_P = 800$, $\mu_R=0.3\mu_P$.  
}
\label{fig4}
\end{figure*}

%=====================
%             Sub-Section  IIIc
%=====================

\subsection{Numerical results}  
\label{sec3c}

We performed simulations for various combinations of the parameters that enter 
inequality (\ref{security_cond}). 
More precisely, for a given pair of $\{\mu_{P},N\}$, we calculated the entropy of the density 
matrix (\ref{rho:eq1}), which is equal to the Holevo bound (\ref{chi:eq}). The estimated bound was used in 
inequality (\ref{fano1:eq}) to obtain the lower bound $p_{\rm err}^{({\rm low})}$ on the error probability. 
The difference of probabilities in inequality (\ref{security_cond}) was estimated numerically  through 
Eqs. (\ref{av_prob3:eq}),  (\ref{theta_av_P:eq}),  and  (\ref{Q_theta:eq}),  
for various combinations of  $\mu_R$, $\bar{\Delta}$ and $\eta$.

The difference $\aver{\hat{Q}_{\widetilde{k}}(\theta)}-\aver{\hat{Q}_k(\theta)}$ enters the inequality 
(\ref{security_cond}) through Eq. (\ref{theta_av_P:eq}), and together with $p_{\rm err}^{({\rm low})}$, 
they determine the security of the protocol for a given $\varepsilon$.  
According to Eq. (\ref{Q_theta:eq}), $\aver{\hat{Q}_{\widetilde{k}}(\theta)}$ and $\aver{\hat{Q}_k(\theta)}$  
have a common amplitude $\sqrt{2\mu_{R}}$ given by (\ref{mu_R:eq}),  and different phases given by 
Eq. (\ref{psi_k:eq}) for $k$ and $\widetilde{k}$, respectively. 
Throughout this analysis we assume that the adversary aims at emulating the verification of the fixed PUK  
on a fixed set-up, using his knowledge on the corresponding CRPs \cite{remark1}. The mean number of scattered photons $\mu_{R}$ that are expected   at the verifier's  HD set-up, is  known to the adversary. Moreover, the adversary knows $\arg({\cal F})$, 
which is  a common fixed phase for both $\aver{\hat{Q}_{\widetilde{k}}(\theta)}$ and 
$\aver{\hat{Q}_{k}(\theta)}$, because it depends only on the set-up and the scattering matrix of 
the PUK which are assumed to be fixed \cite{NikDiaSciRep17}.  Thus, $\arg({\mathscr F})$ is not expected to play any role in 
the inequality (\ref{security_cond}), which has been confirmed by our simulations, and the following 
results and conclusions hold for any value of  $\arg({\cal F})$.  

\begin{figure}
\centerline{\includegraphics[scale=0.42]{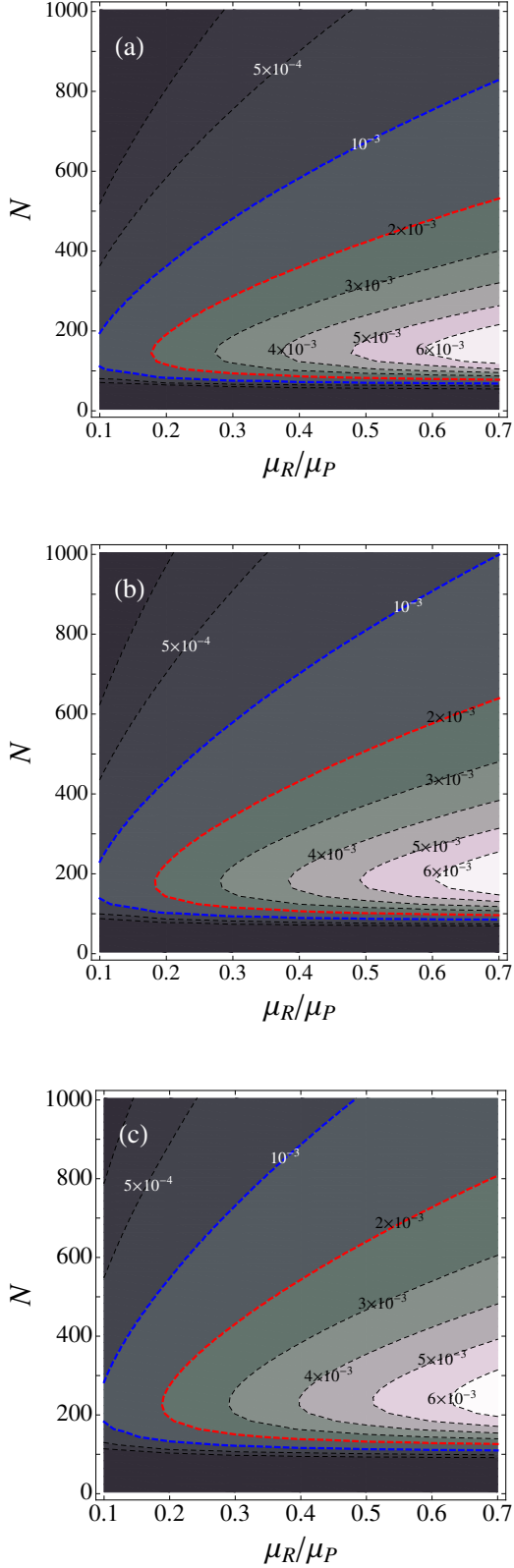}}
\caption{(Color online) The l.h.s of the security condition (\ref{security_cond}) is plotted as a 
function of  $N$ and the ratio $\mu_R/\mu_P$, for $\mu_P = 400$ (a),  $600$ (b) and $1000$ (c). 
For given $\varepsilon$, the security condition is satisfied  in the area enclosed by the contour $D=2\varepsilon$. The contours for various values of $\varepsilon \in [1.5\times 10^{-4},\, 3\times 10^{-3}]$ are shown. Other parameters: $\eta = 0.5$, $\bar{\Delta} = 0.2$. 
}
\label{fig5}
\end{figure}

Our main findings are presented in Figs. \ref{fig4} and \ref{fig5}.  For all of the combinations of parameters, $D$ exhibits an asymmetric bell-like behavior as a function of $N$. 
In particular,  there is a critical value of $N$, say $N_{\rm c}$,  which depends strongly  on $\mu_R$, and  weakly on $\eta$ and on $\bar{\Delta}$.  For $N<N_{\rm c}$, $D$ increases with increasing $N$, while it decreases for 
 $N>N_{\rm c}$. This behaviour can be explained by means of Fig. \ref{fig3}, and noting that $D$ is a product of 
 two terms.  For small values of $N$ we have $D\simeq 0$,  because there is practically no overlap between the possible probe states, and the cheater can discriminate 
 between them (i.e., $p_{\rm err}^{(\rm low)}\simeq 0$). If we keep increasing $N$ for fixed $\mu_P$, 
 the probe states are coming 
closer and they start overlap. Hence, $p_{\rm err}^{(\rm low)}$ becomes non-zero, and increases as we increase 
$N$. At the same time, however,  the difference of the probabilities on the l.h.s. of (\ref{security_cond}) 
decreases, because for any of the quadratures, the photocount distributions associated 
with neighbouring probes are coming closer. Hence, the origin of an 
outcome that falls in the expected bin becomes less certain 
(i.e., the sample may have been obtained from the expected distribution, or   from one of its neighbouring distributions). 
In other words, the server's ability to discriminate between different distributions by looking only at the binned data is gradually 
lost as we increase $N$. Given that $D$ is a product of two functions with opposite monotonicity, we 
have the bell-like behaviour depicted in Figs. \ref{fig4} and \ref{fig5}. 

For typical HD set-ups $\eta\simeq 0.65$, but for the sake of generality we have performed simulations 
for various values of  $\eta$, with $0.5\leq \eta< 1$. Moreover, we consider $4\leq N\leq 10^3$, which is 
within reach of the currently available phase modulators (typical precision 0.3-3 mrads). 
As depicted in Fig. \ref{fig4}(a)-(b), for fixed $\varepsilon$,  the range of values of $N$ for 
which $D$ exceeds $2\varepsilon$ becomes narrower as we decrease $\eta$, while the maximum value 
$D_{\rm max}$ also decreases. 
However, in practise the detection efficiency is limited by the available technology.  
Hence, it is sufficient for our purposes to consider the worst-case scenario for the detection efficiency 
$\eta$, which in our case is $\eta=0.5$. On the contrary,  
$\bar{\Delta}$ can be chosen as large as possible relative to $\sigma$, because it pertains to classical processing of data i.e., the binning of the outcomes from the HDs in $M$ sessions.  Following the same reasoning as in the case of $\eta$, the optimal value of $\bar{\Delta}$ is the one for which the curve $D(N)$ is as broad and as high as possible. Our simulations suggest that the optimal value is $\bar{\Delta}_{\rm opt} \simeq 2$ [see Figs. \ref{fig4}(c)-(d)].

Let us consider now the dependence of $D$ on $\mu_R$ and $N$. 
In view of the findings above, in Fig. \ref{fig5} we plot $D$ as a function of $N$ and $\mu_R$, for $\eta=0.5$ and $\bar{\Delta}=2$, 
and different values of $\mu_P$. For the reasons explained in Sec. \ref{sec2a}, we have focused on 
$\mu_R\leq 0.7 \mu_P$.  For a given value of the error $\varepsilon$,  each contour refers to combinations of $N$ and $\mu_R/\mu_P$ for which $D = 2\varepsilon$, and the security condition (\ref{security_cond}) is satisfied in the enclosed area. For a fixed $\mu_P$, the range of values for $N$ and $\mu_R$ for 
which the security condition is satisfied, becomes broader for decreasing 
$\varepsilon$ (e.g., compare the enclosed areas for different contours in the same plot). 
The same holds for fixed $\varepsilon$ and increasing $\mu_P$ (e.g., compare the enclosed areas for the same contour in different plots). 
  
Of course, for practical reasons, one cannot consider arbitrarily small values of $\varepsilon$, 
because according to (\ref{Mth:eq}) the required sample size increases as $\sim \varepsilon^{-2}$. 
As discussed in Ref. \cite{NikDiaSciRep17}, verification tests of error $5\times 10^{-4}  \lesssim 
\varepsilon \lesssim 1\times 10^{-3}$ and confidence 99.9\% are within reach of today's technology, and ensure security of the protocol in a tamper-resistant scenario. Hence, the contours for $10^{-3}$ and $2\times 10^{-3}$, are of particular interest, in the sense that they are associated with sample sizes for which the protocol is both practical and cloning-resistant. In both cases, we find that the enclosed (secure) areas are rather broad, which suggests that in practise one has a lot of freedom in choosing $N$ and $\mu_P$, so that the protocol is secure, in spite of the details of the implementation, provided that $\mu_R\gtrsim 0.1 \mu_P$. 

In closing, it is worth emphasizing that in the absence of related implementations, 
our simulations have covered a broad range of ratios $\mu_R/\mu_P$. 
In practice, this ratio is expected to be fixed by the verification set-up, and it can be estimated easily 
by means of related measurements. 
Hence, using condition (\ref{security_cond}), one can estimate combinations of 
$\mu_P$, $N$ and $\varepsilon$, such that the protocol is secure against the emulation attack under consideration.

%%%%%%%%%%%%%%%%%%%%%%%%%
%             SECTION  V
%%%%%%%%%%%%%%%%%%%%%%%%%

\section{Discussion}  
\label{sec4} 

Emulation attack is a very general and powerful attack against EAPs that rely on challenge-response mechanisms, and aims at the impersonation of a legitimate user. Generally speaking, the details of the emulation attack depend strongly on the EAP to be attacked, but the framework is always the same. That is, the verification set-up is not tamper-resistant, and the adversary knows the PIN, and the challenge-response properties of the user's key (conventional chip-based smart card or PUK) \cite{remark5}. Hence, the adversary's task is to intercept each challenge during the verification stage, read it, and send to the verifier the expected response. Perhaps the most important feature of the emulation attack is that it can be adapted to any EAP, classical or quantum. E
APs that are used for everyday transactions at automatic teller machines, 
as well as all of  the EAPs that rely on classical read-out of optical  \cite{Horstmeyer15,Tuyls05,Pappu02, PappuPhD,Buch05,Iesl16} or electronic PUKs \cite{e-puks}, are susceptible to an emulation attack. 

In this work we have we have analysed the security of  the EAP of Ref. \cite{NikDiaSciRep17} against an emulation  attack, under the assumption   of identical enrolment and verification set-ups. The adversary 
knows the challenge-response properties of the user's PUK and the related PIN, but he does not have access to the laser source, to the detection set-up of the verifier, and to the random choice of the challenges during the verification. Hence, the most likely attack that he can launch is the emulation attack. More precisely, his task is to intercept each quantum challenge during the verification stage, measure it, and send to the verifier's detection set-up the corresponding response. Our analysis is rather general and does not involve any assumptions about the type of the measurement. Using Holevo's bound and Fano's inequality,  we derived a sufficient condition for an 
implementation of the protocol to be secure against the emulation 
attack. The present results suggest that the protocol of 
Ref. \cite{NikDiaSciRep17} can be  practical, and  simultaneously 
secure against cloning and emulation attack. 

For the time being it is not known whether the emulation attack is the most powerful attack against our EAP, under the aforementioned assumptions. As discussed above, the emulation attack is certainly a powerful and general attack, which can break any EAP that relies on classical read-out of optical or electronic PUKs.  
By contrast, our EAP is robust against the emulation attack, because the PUK is read-out by means of randomly chosen non-orthogonal quantum states. 

The security of the present protocol against other types of attacks, in which the adversary has access to various components of the verification set-up (e.g. to the laser source, or the detection device) goes beyond the present theoretical framework, and it is a subject of future theoretical and experimental work.  Moreover, the performance of the protocol when the verification set-up deviates from the enrolment set-up  (with respect to losses, imperfections, etc), is an interesting question that deserves a thorough investigation (see appendix \ref{app1}). Such deviations are expected to allow for new types of attacks, including perhaps more sophisticated versions of the attack considered in this work. 

The emulation attack considered throughout this work can be implemented with today's technology, because it only requires quantum measurements on coherent states of light. An adversary, however,  may also attack our EAP using a quantum computer 
or a large-scale photonic simulator. Related security analysis for the EAP 
of Ref. \cite{Goorden14} suggests that such attacks require simultaneous 
perfect control of a very large number  of quantum states, or optical elements, with negligible losses, which seems to be 
a formidable challenge for current as well as for future technology. Analogous arguments are expected to be applicable to our scheme as well. 

To summarize, quantum-optical EAPs is a very new field of research, with only few relevant publications \cite{Goorden14,NikDiaSciRep17}. 
There are many open questions about the power and the limitations of such protocols, as well as about the different kinds of attacks 
and their classification. 
The present results provide a benchmark case and guide for the planning of future experiments on the protocol of Ref. \cite{NikDiaSciRep17}.

%\section {Acknowledgements} 

%%%%%%%%%%%%%%%%%%%%%%%%%
%                APPENDIX  
%%%%%%%%%%%%%%%%%%%%%%%%%

\appendix

\section{Losses}
\label{app1}

The key component of our EAP is the interrogation chamber, which is a standard wavefront-shaping set-up, and is connected to the laser source and the HD set-up via SMFs A and B, respectively \cite{NikDiaSciRep17}. 
In the diffusive limit, the operation of the wavefront-shaping set-up can be described in terms of a finite number of transverse spatial modes, 
say  ${\cal N}$ input and  ${\cal N}$ output modes.  Adopting an input-output formalism, the mode of SMF A couples to each one of the ${\cal N}$ input modes, whereas SMF B is translated at the output plane and couples only to the target mode \cite{NikDiaSciRep17}.  The directional transfer of the scattered photons to the target mode is never complete, and part of the input light will be inevitably scattered to the other output modes. Hence, the amount of control the wavefront shaping offers over the propagation of light is quantified by the enhancement factor ${\cal E}$, which in practise it does not exceed the ideal value of ${\cal \pi}{\cal N}/4$. In addition, to the imperfect concentration of scattered light at the target mode, one expects losses which can be viewed as coupling of the light to other modes, beyond  the $2{\cal N}$ input/output modes of the wavefront shaping set-up, and the two SMFs. 

To account for losses in the interrogation chamber and in the SMFs, we can introduce optical power transmissions $\tau_{IC}$, and $\tau_{A(B)}$, respectively.  The mean number of photons at the exit of SMF A is 
\bea
\mu_A = \tau_A \mu_{P},
\label{muA}
\eea
where $\mu_P$ is the mean number of photons in the probe state at the entrance of SMF A.  The mean number of scattered photons at the entrance of SMF B in the case of optimized SLM  is given by \cite{NikDiaSciRep17}
\bea
\mu_{B} =\tau_{IC}  {\cal E} 
\left [
\frac{1}{\cal N}\left (1-\frac{l}L\right )\right ]
\mu_{A}.
\label{muIC}
\eea
In deriving this expression we have assumed slab geometry for the PUK, with thickness $L$, mean free path $l$, and absorption length $L_{\rm abs}$, such that  $\lambda\ll l\ll L \ll L_{\rm abs}$, where $\lambda$ is the wavelength of the probe.
The transmission coefficient $\tau_{IC}\lesssim 1$ accounts for losses in the interrogation chamber, including  absorption losses, imperfect couplings etc. The mean number of photons that reach the HD set-up 
is 
\bea
\mu_{R} = \tau_B \mu_B.
\label{muR}
\eea 
Combining Eqs. (\ref{muA}) -  (\ref{muR}) we obtain 
\bea
\mu_R={\cal E} |{\cal F}|^2 \mu_P
\eea 
where
\bea
|{\cal F}|^2 =  \tau_B\tau_{IC}   \tau_A
\left [
\frac{1}{\cal N}\left (1-\frac{l}L\right )\right ].
\eea
We see therefore that $|{\cal F}|^2$, and thus $\mu_{R}$, are  independent of $k$, and include all of the losses throughout the propagation of the light in the set-up. 

In the security analysis of the present work we assume identical enrolment and verification set-ups. All of the parameters of the set-ups (including losses, imperfections, enhancement, etc), are publicly available (see table \ref{tab2}). The mean number of photons  that are expected to reach the HD set-up of the verifier, i.e., $\mu_{\rm R}$, is also publicly known.  
The security of the EAP relies solely on keeping secret the random,  independently chosen phases of the probe states. Acceptance or rejection of the PUK is decided upon the estimated average probability for an outcome to fall in the bin, which is also publicly known. 

If the adversary tries to reduce or increase the losses in the set-up, relative to the publicly known values, the verifier will inevitably receive responses that differ from the expected ones both in amplitude (mean number of photons), and in phase. The security analysis of Sec. \ref{sec3} can be adapted to this case as well,  by taking into account that $\langle \hat{Q}_{\widetilde{k}}(\theta) \rangle$ will differ  from $\langle \hat{Q}_k(\theta) \rangle$, both in phase and in amplitude.   In analogy to the discussion of Sec. \ref{sec3}, the adversary's intervention will shift the photocount distributions the verifier samples from, relative to the expected bins. It does not matter whether this shift will be above or below $\langle \hat{Q}_k(\theta) \rangle$, because the r.h.s. of Eq. (\ref{Pexp_cheating:eq}) is an even function of $S$.  

In the envisioned implementation of our EAP,  the three major parts of the verification set-up  i.e.,  the laser source, the interrogation chamber and the detection set-up (see Fig. \ref{fig1}), 
are located in the same or nearby rooms. 
This means that their separation is 
tens of meters, and thus the losses in the optical fibers that connect them 
are negligible. Typically, for optical fibres at 1550 nm, the attenuation coefficient is $\sim 0.2 {\rm dB/km}$, which means almost perfect transmittance at these distances (i.e., $\tau_{A(B)}\gtrsim 0.99$). 
 
In closing this appendix it is worth pointing out that effects of losses and imperfections may play some role in the security of the protocol, if the verification set-up deviates from the enrolment set-up. This scenario goes beyond the present work, and deserves a careful security analysis, because other types of attacks may be applicable in this case.  We strongly believe, however, that for sufficiently small deviations, one can choose the bins' width $\Delta$ judiciously so that the protocol tolerates the deviations.

\section{Binning of normal distribution}
\label{app2}
Consider the normal distribution ${\mathscr N}(\bar{x},\sigma^2)$, with probability density
\bea
{\cal P}(x) = \frac{1}{\sqrt{2\pi\sigma^2}}\exp\left (- \frac{(x-\bar{x})^2}{2\sigma^2}\right )
\eea  
and a bin of size $\Delta$ centred at $x=b$ 
\[
\left [b-\frac{\Delta}2,~b+\frac{\Delta}2 \right ]. 
\]
The probability for a random sample from ${\mathscr N}(\bar{x},\sigma^2)$, to yield an outcome that falls in the 
bin is given by the following integral 
\bea
P_{\rm in} &=& \int_{-\frac{\Delta}2}^{\frac{\Delta}2} dy {\cal P}(y+b) 
\nonumber\\
&=& \frac{1}{2}\left \{
{\rm Erf}\left [ 
\frac{2(\bar{x}-b)+\Delta}{2\sqrt{2}\sigma}
\right ] - {\rm Erf}\left [ 
\frac{2(\bar{x}-b)-\Delta}{2\sqrt{2}\sigma}
\right ]
\right \}
\nonumber\\
&=&\frac{1}{2}\left \{
{\rm Erf}\left [ 
\frac{2\bar{\xi}+\bar{\Delta}}{2\sqrt{2}}
\right ] - {\rm Erf}\left [ 
\frac{2\bar{\xi}-\bar{\Delta}}{2\sqrt{2}}
\right ]
\right \},
\eea
where ${\rm Erf}(\cdot)$ is the error function,  $\xi := \bar{x}-b$, $\bar{\xi} := \xi/\sigma$ and $\bar{\Delta} := \Delta/\sigma$. 

We see that $P_{\rm in}$ depends on the distance  $\xi$ between the centres of the Gaussian and of the bin, 
as well as on the  size of the bin $\Delta$, relative to  the standard deviation $\sigma$.   
When the centers coincide i.e., for $\xi =0$, the above expression simplifies to Eq. (\ref{Pexp:eq}).

\section{Upper bound on probability for in-bin event, in the presence of cheating}
\label{app3}

Consider a query (sample) with coherent probe state $\ket{\alpha_k}$, which is fully characterized by 
the randomly chosen integer  $k\in \Int_N$. 
An adversary intercepts and measures the unknown probe state, so that to learn $k$. 
Let $\widetilde{k}$ denote the adversary's inference, and let $\beta_{\widetilde{k}}$ denote the coherent state that is sent to the 
verifier's HD set-up.    The verifier measures at random one of the two quadratures of the electric field, by setting 
the phase of the LO $\theta$, either to 0, or to $\pi/2$. 

The average probability for the verifier to obtain an outcome in the expected bin is given by 
\bea
P_{\rm in} &=& 
\sum_{k,\widetilde{k},\theta} P({\rm in},k,\widetilde{k},\theta)
=\sum_{k,\widetilde{k},\theta} P({\rm in}|k,\widetilde{k},\theta)p(k,\widetilde{k})p(\theta) 
\nonumber 
\\
&=&\sum_{k,\theta} P({\rm in}|k,\widetilde{k}=k,\theta)p(k,\widetilde{k} = k)p(\theta)
\nonumber \\ 
&&+\sum_{k}\sum_{\widetilde{k}\neq k}\sum_{\theta} 
P({\rm in}|k,\widetilde{k},\theta)p(k,\widetilde{k}) p(\theta) 
\label{Pin_cheat:eq}
\\
&:=& S_1+S_2,
\nonumber
\eea
where we have used the fact that the angle $\theta$ is chosen at random and independently of the choice of $k$, while the adversary does not have access to the preparation of the probe states, or to the HD set-up. 
Hence,  the inference of the right or wrong value of $k$ by the adversary is independent of the 
quadrature to be measured by the verifier.  

Let us consider first the summation $S_1$. 
The conditional probability $P({\rm in}|k,\widetilde{k}=k,\theta)$ is the probability for an outcome to fall in the expected bin, given that the verifier has prepared $\ket{\alpha_k}$, the cheater has inferred the right value of $k$, and the verifier measures the $\theta$ quadrature. Clearly, 
this probability is given by Eq. (\ref{Pexp:eq}) (i.e., it is independent of $k$ and $\theta$), because the cheater's intervention has not changed anything relative to the ideal scenario. The joint probability 
$p(k,\tilde{k} = k)$ is the probability for the verifier to send $k$, and the 
cheater to infer the right value of $k$. Thus, 
\bea
\sum_k p(k,\tilde{k} = k) = 1 -p_{\rm err},
\label{sum1A:eq}
\eea
where $p_{\rm err}$ is the average probability for the adversary to infer a wrong value for $k$.
So, we have 
\bea
S_1 &=&\sum_{k,\theta} P({\rm in}|k,\tilde{k} = k,\theta)p(k,\tilde{k} = k)p(\theta) 
\nonumber \\
&=& 
 P_{\rm in}^{(0)} \sum_{k}p(k,\tilde{k} = k) = (1-p_{\rm err} )  P_{\rm in}^{(0)}. 
 \label{S1:eq}
\eea 

We focus now on the second term in Eq. (\ref{Pin_cheat:eq}). The joint probability 
 $p(k,\widetilde{k})$ is the probability for the verifier to send $k$, and the 
cheater to infer a wrong value for $k$, in the particular case $\widetilde{k}$. Thus, if we sum over all possible values of $k$ and $\widetilde{k} \neq k$ 
we will obtain the average probability of error  
\bea
\sum_k \sum_{\widetilde{k}\neq k} p(k,\widetilde{k} ) = p_{\rm err},
\label{sum2A:eq}
\eea
which also appears in Fano's inequality (\ref{fano1:eq}).
So, we have 
\bea
S_2&=&\sum_{k}\sum_{\widetilde{k}\neq k}\sum_{\theta} 
P({\rm in}|k,\widetilde{k},\theta)p(k,\widetilde{k}) p(\theta) 
\\
&=&
\sum_{k}\sum_{\widetilde{k}\neq k}
p(k,\widetilde{k})\sum_{\theta=0,\pi/2} 
P({\rm in}|k,\widetilde{k},\theta)p(\theta) 
\\
&=&
\sum_{k}\sum_{\widetilde{k}\neq k}
p(k,\widetilde{k}) P({\rm in}|k,\widetilde{k}), 
\eea
where $P({\rm in}|k,\widetilde{k}) $ is the conditional probability for an outcome to fall in the expected bin, given that the verifier has prepared $\ket{\alpha_k}$, the cheater has inferred  $\widetilde{k}\neq k$, irrespective of the  measured quadrature. 

Hence, using Eq. (\ref{sum2A:eq}) we can bound $S_2$ from above as follows
\bea
S_2 \leq  
p_{\rm err} \max_{k,\widetilde{k}}\{P({\rm in}|k,\widetilde{k}) \} _{\widetilde{k}\neq k}.
\label{S2:eq}
\eea

Using (\ref{S1:eq}) and (\ref{S2:eq}) in (\ref{Pin_cheat:eq}) we have 
\bea
P_{\rm in} \leq (1-p_{\rm err})P_{\rm in}^{(0)}+ p_{\rm err} \max_{k,\widetilde{k}}\{\bar{P}({\rm in}|k,\widetilde{k}) \} _{\widetilde{k}\neq k}.
\eea


\begin{thebibliography}{99}
\bibitem{remark1} 
The verification of the PUK takes place only if the server has confirmed the validity of the PIN. If the PIN is not valid, the verification is aborted immediately. Hence, a necessary condition for the emulation attack to take place is the adversary to know the PIN of the user he intends to impersonate, together with the CRPs of the user's PUK.  
One can assume that the database of CRPs is compromised, and the adversary 
has obtained a copy of the set of all the CRPs to be used in the verification of the PUK, together with the PIN. 
Alternatively, one may assume that the adversary has obtained the PIN by other means, and at some point in the past, he  has obtained unnoticed 
access to the PUK for a sufficiently large amount of time, so that to estimate its entire scattering matrix 
before he returns it to the owner. The security analysis in this work applies to both of these scenarios, although for the sake of concreteness, it is the first one that is mainly  mentioned in the  text.
\bibitem{remark4}
The protocol under consideration can be also implemented 
with optimized phase mask that changes with the phase of the probe. The analysis of Sec. \ref{sec3a} and the security condition (\ref{security_cond}) are expected to be valid for this case as well, provided one 
takes into account the dependence of ${\cal E}$ and ${\cal F}$ on $k$, when estimating  
$\max_{k,\tilde{k}}\{P({\rm in}|k,\tilde{k})\}_{k\neq \tilde{k}}$
in condition (\ref{security_cond}).
\bibitem{remark2}
Here we assume that the server chooses the random challenges and the quadratures to be measured, and sends them to the verifier. 
Alternatively, one can assume that the verifier obtains a copy of the entire set of CRPs pertinent to the PUK 
under authentication, and he chooses locally at random and independently the $M$ challenges and the quadratures to be measured. 
In any case, the verifier should never reveal the sequence of random challenges or the measured quadratures, even after the verification is over.
\bibitem{remark6} 
We assume that the server has a reliable random number generator (RNG), for the random choice of  challenges and quadratures to be measured. 
Any flaws in the operation of the RNG  can be, in principle, exploited by adversaries. Quantum random generators of high quality are currently available on the market \cite{qrng}.
\bibitem{remark3}
If the verification device is activated only when a PUK is detected to be located in the corresponding slot, then 
the adversary can insert a false PUK, because he bypasses the interrogation chamber and thus the inserted PUK (true or false), does not play any role in the attack.
\bibitem{remark5}
Clearly, if the adversary has stolen the user's PUK, and moreover knows the PIN, he can impersonate successfully the user. In this case, purely cryptographic identification cannot distinguish between the legitimate user and the adversary. To this end, one has to resort to EAPs that rely on biometrics i.e., on physical characteristics of the human body.
Such EAPs go beyond the scope of the present work. However, it is worth emphasizing that there are many implementation issues, both technical,
practical, and sociological.
\end{thebibliography}
\end{document}